# Towards a Cellular Automata Based Network Intrusion Detection System with Power Level Metric in Wireless Adhoc Networks (IDFADNWCA)

P.Kiran Sree [1], I. Ramesh Babu [2]

[1] Research Scholar, Dept of C.S.E, Jawaharlal Nehru Technological University, Hyderabad.
[2] Professor, Dept of C.S.E, Acharya Nagarjuna University, Guntur.
Email: [1] profkiran@yahoo.com, [2] rinampudi@hotmail.com

## Abstract

*Adhoc wireless network with their changing topology and distributed nature are more prone to intruders. The efficiency of an Intrusion detection system in the case of an adhoc network is not only determined by its dynamicity in monitoring but also in its flexibility in utilizing the available power in each of its nodes. In this paper we propose a hybrid intrusion detection system, based on a power level metric for potential adhoc hosts, which is used to determine the duration for which a particular node can support a network-monitoring node. The detection of intrusions in the network is done with the help of Cellular Automata (CA). IDFADNWCA (Intrusion Detection for Adhoc Networks with Cellular Automata) focuses on the available power level in each of the nodes and determines the network monitors. Power Level Metric in the network results in maintaining power for network monitoring, with monitors changing often, since it is an iterative power optimal solution to identify nodes for distributed agent based intrusion detection. The advantage of this approach entails is the inherent flexibility it provides, by means of considering only fewer nodes for reestablishing network monitors.*

## 1. Introduction

An intrusion is defined as any set of actions that attempt to compromise the integrity, confidentiality, or availability of a resource. Intrusions in wireless networks amount to interception, interruption, or fabrication of data transmitted across nodes, wherein an intruder node attempts to access unauthorized data. Intrusion detection is one of key techniques behind protecting a network against intruders. An Intrusion Detection System is a system that tries to detect and alert on attempted intrusions into a system or network, where an intrusion is considered to be any unauthorized or unwanted activity on that system or network [2]. Adhoc networks are particularly prone to such dangers, considering the dynamic and geographically distributed nature of the nodes.

## 2. Survey of Related Work

Numerous detection systems have been proposed to tackle the problem of intrusion in wireless networks some of which are an extension of intrusion detection system in wired networks. Few deal with network based IDS [6] and few with host based IDS [4], all which are based on lightweight agents [5, 6]. Power awareness in mobile adhoc networks [4] becomes a major issue when considering intrusion detection in larger networks.

### 2.1. Triggering Mechanisms

To protect the network, IDS must generate alarms when it detects intrusive activity on the network. Different IDS trigger alarms based on different types of network activity.
The two most common triggering mechanisms are the following:
1. Anomaly detection
2. Misuse detection

Discussions regarding the above triggering mechanisms can be found in [3]. Besides implementing a triggering mechanism, the IDS must somehow watch for intrusive activity at specific points within the network. Monitoring intrusive activity normally occurs at the following two locations:
1. Host – Host based IDS

2. Network – Network based IDS

Finally, many intrusion detection systems incorporate multiple features into a single system. These systems are known as hybrid systems. These hybrid intrusion detection systems having their architecture based on agents [2, 6] which travel throughout the network, provide a comprehensive solution.

## 3. IDFADNWCA & Cellular Automata (CA)

Cellular automata use localized structures to solve problems in an evolutionary way. CA often demonstrates also significant ability toward self-organization that comes mostly from the localized structure on which they operate. By organization, one means that after some time in the evolutionary process, the system exhibits more or less stable localized structures [4]. This behavior can be found no matter the initial conditions of the automaton.

A CA consists of a number of cells organized in the form of a lattice. It evolves in discrete space and time. The next state of a cell depends on its own state and the states of its neighboring cells. In a 3-neighborhood dependency, the next state $q_i(t+1)$ of a cell is assumed to be dependent only on itself and on its two neighbors (left and right) and is denoted as:

$$q_i(t+1) = f(q_{i-1}(t), q_i(t), q_{i+1}(t)) \qquad (1)$$

where, $q_i(t)$ represents the state of the $i^{th}$ cell at $t^{th}$ instant of time, f is the next state function and referred to as the rule of the automata. The decimal equivalent of the next state function, as introduced by Wolfram, is the rule number of the CA cell.

### 3.1 Fuzzy CA fundamentals

FCA is a linear array of cells which evolves in time. Each cell of the array assumes a state qi, a rational value in the interval [0,1] (fuzzy states) and changes its state according to a local evolution function on its own state and the states of its two neighbors. The degree to which a cell is in fuzzy states 1 and 0 can be calculated with the membership functions. This gives more accuracy in intrusion. In a FCA, the conventional Boolean functions are AND, OR, NOT.

### 3.2 Dependency matrix for FCA

Rules defined in Equation. 1 should be represented as a local transition function of FCA cell. That rules are converted into matrix form for easier representation of chromosomes(Figure 1). Rules will be used for accommodating dynamism into the network.

$$T = \begin{bmatrix} 1 & 1 & 0 & 0 \\ 1 & 1 & 1 & 0 \\ 0 & 0 & 1 & 1 \\ 0 & 0 & 1 & 1 \end{bmatrix}$$

**Figure 1.** Matrix Representation of Rule

### 3.3 Genetic Algorithm & CA

The main motivation behind the evolving cellular automata framework is to understand how genetic algorithms evolve cellular automata that perform computational tasks requiring global information processing, since the individual cells in a CA can communicate only locally without the existence of a central control the GA has to evolve CA that exhibit higher-level emergent behavior in order to perform this global information processing. Thus this framework provides an approach to studying how evolution can create dynamical systems in which the interactions of simple components with local information storage and communication give rise to coordinated global information processing.

### 3.4 Crossover and Mutation

Traditional genetic algorithms have been used to identify and converge populations of candidate hypotheses to a single global optimum. For this problem, a set of rules is needed as a basis for the IDS. There is no way to clearly identity whether a network connection is normal or anomalous just using one rule. Multiple rules are needed to identify unrelated anomalies, which mean that several good rules are more effective than a single best rule. Another reason for finding multiple rules is that because there are so many network connection possibilities.

The performance of an individual is measured by a fitness function. A fitness function rates the performance of an individual in its environment by comparing the results of the individual's chromosomes with the desired results of the problem as defined by the author of the algorithm. The fitness is generally expressed within the algorithm as a floating point (i.e., decimal) number with a predefined range of values, from best performing to worst performing. As in Darwinian evolution, low-performing individuals are eliminated from the population and high performing individuals are cloned and mutated, replacing those that were eliminated.

## 3.5 Basic Algorithm

CA Tree Building (Assuming K[1] CA Basins)

Input:    Intrusion parameters
Output: CA Based inverted tree
Step 0: Start.
Step 1: Generate a CA with *k* number of CA basins
Step 2: Distribute the parameters into k CA basins
Step 3: Evaluate the distribution in each closet basin
Step 4: Calculate the RI (Power Level Metric, Fourier Index, Laplase Transformation)
Step 5: Swap the more appropriate features to the bottom leaves of the inverted tree.
Step 6: Stop.

The main feature which is achieved when developing CA Agent systems, if they work, is flexibility, since a CA Agent system can be added to, modified and reconstructed, without the need for detailed rewriting of the application. These systems also tend to be rapidly self-recovering and failure proof, usually due to the heavy redundancy of components and the self managed.

The agent-based model proposed in [4] approaches the IDS problem with a technique that handles intrusions with an agent running on each system. Further, the model in [4] is not suitable for a power-aware IDS , since such a system warrants energy consumption in systems irrespective of their current battery levels, i.e. it suggests an IDS without considering the feasibility of the assumption that network monitoring and analysis is justified in nodes with minimal power, such as robust wireless sensor networks (WSN).

## 4. IDFADNWCA

The agent-based model proposed in [4] approaches the IDS problem with a technique that handles intrusions with an agent running on each system. Further, the model in [4] is not suitable for a power-aware IDS , since such a system warrants energy consumption in systems irrespective of their current battery levels, i.e. it suggests an IDS without considering the feasibility of the assumption that network monitoring and analysis is justified in nodes with minimal power, such as robust wireless sensor networks (WSN).

## 4.1 Modular IDS Architecture

The IDS we consider is built on a mobile agent framework as in [1]. It is a non-monolithic system and employs several sensor agents that perform certain functions, such as:

A hierarchy of agents has been devised in order to achieve the above goals. We will adapt the hierarchy for our purposes. There are three major agent classes as used in [2], categorized as monitoring, decision-making and action agents. Some are present on all mobile hosts, while others are distributed to only a select group of nodes, as discussed further. The monitoring agent class consists of packet, user, and system monitoring agents. The following diagram (Fig

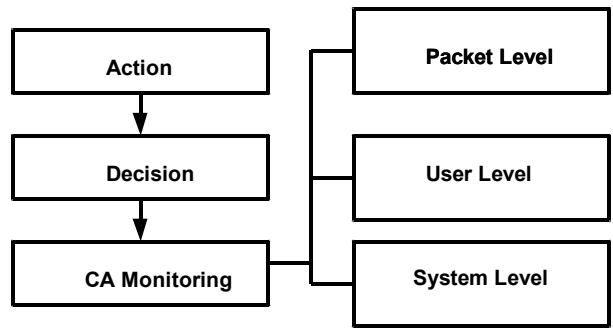

**Figure 2.** Typical agent hierarchy, depicting the multi-level decision making process for intrusion detection

2) shows the hierarchy of agent classes.

## 4.2. The IDFADNWCA Algorithm

In IDFADNWCA, we deal with multi-hop network monitoring clustered node selection, similar to SPAID. This type of a node selection has its inherent advantages in allowing complete coverage of all nodes and links in a network, but with an added factor of redundancy in the collection of intrusion detection data. The algorithm presented here is a power efficient variation over the previous SPAID, where the increment in hop radius and re-running of SPAID was considered for the whole topology, after a particular drop in power for certain monitoring nodes.

This approach considers each of the initially allocated monitors and the nodes they monitor to be a single tree, with the monitoring node as a root and the nodes being monitored as its child. The root along with its child nodes contribute to individual CA clusters.

Thus the whole large topology gets divided to tree structured CA clusters only for intrusion detection purpose. Thus once a node is selected and allotted as a monitor initially for a set of nodes, they form an individual cluster.

After such CA clusters are established, when any drain in power levels take place to the monitors, any other child node with higher battery power for monitoring can take charge of that cluster. Since only certain CA clusters are active for certain period of time, it is enough if their root nodes i.e., their monitoring nodes get re-arranged (within the cluster), instead of running the whole node selection process as in the case of SPAID which is inefficient.

The node selection process should be done for a network as a whole only when no single node within a cluster is competent enough to monitor or when a new node with higher CA PARAMETER value enters the existing network.

The IDFADNWCA algorithm uses the agent hierarchy, with a significantly adapted node selection mechanism to incorporate power-awareness, and is best detailed by the following eight steps.

***Step 1****: Set CA parameter threshold*. Set a constraint on the CA PARAMETER value of nodes which are allowed to compete for becoming a network monitoring node using 3.5 algorithm.

***Step 2****: CA parameter Calculation and CA parameter Ordered List (POL)*. Arrange the different nodes in increasing values of CA parameter as calculated previously with CA basin Values, for all nodes which satisfy the CA parameter Constraint.

***Step 3****: Hop Radius*. Set the hop radius to one initially, and increment for each insufficient node selection with the current hop radius with power level metric.

***Step 4****: Expand Working Set of Nodes*. Consider node selection incrementally, initially from the first node, to finally the set of all nodes in the network, incrementing the set of nodes under consideration by one node each time. We call this set the working set (WS) of nodes. The WS is expanded only if the addition leads to an increase in number of represented nodes.

***Step 5****: Voting. The voting scheme* for Node Selection, is similar to that in [2], except that we limit the candidates to just the nodes which are part of WS.

***Step 6****: Check acceptability of nodes*. If all links/nodes are not represented by the set of nodes covered by the voting scheme, then we expand the WS and repeat the process from Step 4. If WS equals the POL, then increment the hop radius, and repeat from Step 3. It is suggested that the increment in hop radius be considered a final resort, as it effectively increases the amount of processing per monitoring node.

***Step 7****: Cellular Cluster Setup*. Set individual CA clusters with the nodes in the working set as root and the nodes being monitored by it as child nodes.

***Step 8****:   Re-run with CA*. Changes in power levels of the root nodes in each cluster will be signaled to the child and the vote count as in step 5 takes place within the cluster to form a new monitoring node.

### 4.3 Details of the Algorithm

There are two phases to the implementation of classification problem using CA's. They are the Training or Learning phase and Testing or Recognition phase. It has the two phases-training phase and testing.

#### 4.3.1 Training or Learning phase

In the training phase the optimal hyper CA Parameters for each of the binary classifiers are constructed based on the training set data. The system is trained with a lot of sample intrusions and their attributes. These images form the basis in identifying the image the user queries. The attribute vector for each of these intrusions is found and is stored in a feature database.

#### 4.3.2 Testing or Recognition phase

In the testing phase the attributes of the instance are used to query. The attribute vector is calculated for the image and is given as input to the pool of trained CA's which identifies the class to which the instance belongs and takes the necessary steps.

Presently, the work has been conducted only on some sample networks. In other words, the work consists of preliminary findings which offer a base for further expansion and exploration in the study of application of Cellular Automata to Adhoc Networks.

## 5. Experimental Results

Evaluating the extended algorithm, IDFADNWCA, in terms of power, results in much better utilization of the available power (Figure 3). Splitting up of larger networks to CA clusters and manipulating power levels and thresholds within them provides a power optimal solution than that of SPAID, which requires the entire network. Also SPAID [1] was considered only for minimally mobile networks with increments in hop radius for more dense networks, while IDFADNWCA with tree based CA clusters can prove efficient even in the case of dynamic networks.

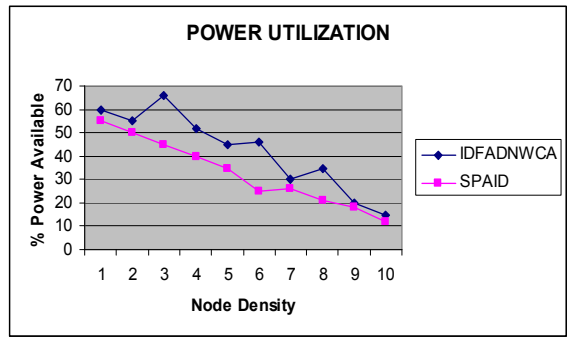

**Figure 3.** Performance Comparison – SPAID vs. IDFADNWCA

## 6. Conclusion

In this paper, sufficient base has been provided to understand the efficient functioning of the IDFADNWCA Algorithm in determining the duration for which a node can support a network monitoring function in wireless adhoc networks. The preliminary results show that the IDFADNWCA Algorithm gives good results on sparse as well as dense mobile networks.

It is evident from the power utilization performance evaluation the IDFADNWCA algorithm proves to be scalable and even more efficient as the number of nodes increases i.e. as the size of the wireless adhoc network increases. Re-run over the entire network for node selection needs to be done often only with changing network topologies.

## 7. References

[1]T. Srinivasan, Jayesh Seshadri, J.B. Siddharth Jonathan,and Arvind Chandrasekhar,"A System for Power-Aware Agent-Based Intrusion Detection (SPAID) in Wireless Ad Hoc Networks", Springer-Verlag Berlin Heidelberg 2005.

[2]Kachirski, O. and Guha, R," Efficient Intrusion Detection using Multiple Sensors in Wireless Ad Hoc Networks", 36th Annual Hawaii International Conference on System Sciences (HICSS'03) ,pp 25-30 , January 06 - 09, 2003.

[3]Zhou, L., and Haas, Z.J.,"Securing Ad Hoc Networks", IEEE Networks Special Issue on Network Security. Pp 30-35, November,1999.

[4]Ahmed Safwat et al, Handbook of Ad hoc Wireless Networks, CRC Press, Dec. 2002, 'Power-Aware Wireless Mobile Ad hoc Networks',

[5]D. Dasgupta and H. Brian, "Mobile Security Agents for Network Traffic Analysis", Proceedings of DARPA Information Survivability Conference & Exposition II, 2001. DISCEX '01, Volume: 2, 2001, pp. 332–340.

[5]G. Helmer, J. Wong, V. Honavar, L, Miller, "Lightweight Agents for Intrusion Detection", Technical Report, Dept. of Computer Science, Iowa State University, 2000.

[6]Chao Gui and Prasant Mohapatra," Power conservation and quality of surveillance in target tracking sensor networks", In Proceedings, MobiCom '04, pages 129.143, New York, NY, USA, 2004. ACM Press.

[7]P. Kiran Sree al.,"Power-Aware Hybrid Intrusion Detection System Using Support Vector Machine in Wireless Ad Hoc Networks", Proceedings of Knowledge Management International Conference,10-12, June, 2008,Langkawi, Malaysia,pp:547-553,

[8]P.Kiran Sree, al. "Improving Quality of Clustering using Cellular Automata for Information retrieval" Journal of Computer Science 4 (2): 167-171, 2008.

[9]P. Maji al," FMACA: a fuzzy Cellular Automata based classifier ", in the proceeding of Ninth International Conference on Database Systems, Korea, pp 494-505